# Calculating Method of Moments Uniform Bin Width Histograms

James S. Weber, Ph.D.
June 5, 2016

**Abstract**. A clear articulation of Method of Moments (MOM) Histograms is instructive and has waited 121 years since 1895. Also of interest are enabling uniform bin width (UBW) shape level sets. Mean-variance MOM uniform bin width frequency and density histograms are not unique, however ranking them by histogram skewness compared to data skewness helps. Although theoretical issues rarely take second place to calculations, here calculations based on shape level sets are central and challenge uncritically accepted practice. Complete understanding requires familiarity with histogram shape level sets and arithmetic progressions in the data.

---------------------------------------------------

## 1. Introduction.

Widely familiar uniform bin width (UBW) histograms are data graphics and density estimators based on bin counts, $v_k$, of data points, $x_i$, $i = 1$ to $n$, in half-open uniform width bins, $[t_o + (k - 1)h, t_o + kh)$, $k = 1$ to $K$. Respectively $t_o$, $h$, $K$ are a bin location anchor point, bin width, maximum number of bins. Frequency, relative frequency and density histogram shape are, respectively, a list $(v_k)$, $(v_k/n)$, or $(v_k/nh)$, $k = 1$ to $K$. Histogram *shape is pivotal* in obtaining many kinds of optimal histograms. Shapes and shape level sets lead to *local* optima which lead to global optima.

Although the earliest known use of histogram grouped data is 1662, possibly not until 1895 did they have a prominent advocate, Karl Pearson (1895). Karl Pearson also advocated Method of moments (MOM) estimation (1894, 1895, 1902). In its simplest form model parameter estimates are calculated so that selected model moments and data moments are the same. (MOM can include a system of distribution curves although clearly MOM can be done simply as here, or more generally, e.g. Hansen, L. P. 1982.)

Why haven't MOM histograms been explained until now? Mean and variance constraints (1ab), (2ab), §2.2 on bin parameters ($t_o$, $h$) for frequency and density histograms are easy to understand. The following troubles and misperceptions may explain why.

1. For some shapes, obtaining $t_o$ and $h$ from bin counts, $v_k$, with mean and variance constraints, (1b), (2b), §2.2, leads to bins that lead to different bin counts. That is, the MOM estimates are outside of a relevant ($t_o$, $h$) parameter space for bin counts, $v_k$.

2. As a consequence of #1, there could be hope for a *unique shape* that satisfies mean and variance moment constraints. However *many shapes* associate with ($t_o$, $h$) values leading to the same bin counts and frequency histogram grouped data mean and variance, or histogram density mean and variance that agree with data mean and variance.

3. It might be expected that histograms agreeing with the data mean and variance must have similar shapes. But this is not true. Great variety of shapes for small samples may not be fully appreciated. Tables 1ab, col $J_g$, §2.5, show this.

4. For histograms, MOM mathematics is elementary, resonating with R. J. Little (2013). "In Praise of Simplicity not Mathematistry! Ten Simple Powerful Ideas for the Statistical Scientist" *JASA*. There is meager return on investment, there is little incentive for solving *apparently difficult* or *relatively complicated elementary problems*.

This is different from familiar applications of elementary MOM estimation, such as MOM estimation of normal distribution parameters. Further, research on histograms since Sturges (1926) appears to have focused primarily on number of bins and bin width. Although bin location often is varied in the calculation of histogram estimates, location often is absent in asymptotic optimal histogram results. However understanding and calculating MOM histograms, especially for smaller samples, depends critically on location through shape level sets. *Shape level sets* (Appendix A) *unify mutual dependence of bin location, bin width, histogram shape and data*.

**2.1 UBW Histogram Shapes, Level Sets, Shape and Moment Consistency**

**Definition**. **Histogram shape** is a list of bin counts, ($v_k$); or relative frequencies, ($v_k/n$); or density *step function* values, ($v_k/nh$), $k = 1$ to **K**.

**Definition**. For fixed data, UBW **Histogram shape level sets** are sets of ($t_o$, $h$) values that lead to bins that lead to the same shapes: ($v_k$), ($v_k/n$), ($v_k/nh$), $k = 1$ to **K**. Shape level sets are convex polygons in {($t_o$, $h$)} and are identical for frequency, relative frequency and density histograms. (Appendix A, especially Figure A)

**Definition**. **Histogram shape & data-moment consistency**. A histogram *shape* is consistent with a data moment if there is/are bin location(s) and width(s) leading to the same shape and histogram grouped data or density moment agreeing with a data moment. (I.e. MOM parameter estimate(s) are in a shape level set parameter space.) Focusing on moments one at a time usually under identifies, but is the only way to see the whole picture. For frequency and density histograms, MOM parameter estimators are *identical for mean and skewness, but <u>not</u> variance*. Density histogram variance = (frequency histogram grouped data variance) + ($h^2/12$) = (*between bin* variance) + (*within bin* variance).

**2.2 Frequency histogram mean, variance and skewness constraints.** Calculating histogram *grouped data* moments replaces data values with bin midpoints, ($t_o + (k - ½)h$), with bin frequencies $v_k$. (Histogram density moments are moments of step function histogram densities.) Expressions (1a)-(3e), below, and Appendix D, review grouped data mean, variance, skewness, etc. Subscripts "g" and "x" distinguish grouped data and sample statistics. (1b) and (2b) show bin width, $h$, as a straight line in $\{(t_o, h)\}$ in terms of other variables.

$$\textbf{Grouped data } \textit{mean} \equiv \bar{x}_g \equiv \frac{1}{n}\sum_{k=1}^{K} v_k [t_0 + (k - \frac{1}{2})h] =$$

$$= t_o - \frac{h}{2} + \frac{h}{n}\sum_{k=1}^{K} k v_k = t_0 + h(\bar{k} - \frac{1}{2}) = \bar{x}_x, \quad (1a)$$

$$h = \bar{x}/(\bar{k} - \frac{1}{2}) - t_0/(\bar{k} - \frac{1}{2}); \quad x_{min} + h < t_0 \leq x_{min} \quad (1b)$$

$$\textbf{Grouped data } \textit{variance} \equiv s^2_g = \frac{h^2}{n-1}[\sum_{k=1}^{K} v_k(k-\bar{k})^2] = s^2_x, \quad (2a)$$

$$h = [(n-1)s^2_x / \sum_{k=1}^{K} v_k(k-\bar{k})^2]^{½} + 0\, t_o \quad (2b)$$

Implementations of skewness include grouped data and sample gamma, $g_g$, $g_x$, (3a), (3b); and Fisher-Pearson adjusted third moment coefficients, $FPS_g$, $FPS_x$, (3c), (3d) found in Excel©® and statistical packages including Minitab©®, SAS©®, SPSS©®.

$$\textbf{Grouped data } \textit{gamma skewness} \equiv g_g \equiv \frac{\frac{1}{n}\sum_{k=1}^{K} v_k(k-\bar{k})^3}{(\frac{1}{n}\sum_{k=1}^{K} v_k(k-\bar{k})^2)^{3/2}} = f(v_k, K, n) \sim= g_x \quad (3a)$$

$$\textbf{Sample } \textit{gamma skewness} \equiv g_x \equiv \frac{\frac{1}{n}\sum_{i=1}^{n}(x_i - \bar{x})^3}{(\frac{1}{n}\sum_{i=1}^{n}(x_i - \bar{x})^2)^{3/2}} \quad (3b)$$

**Grouped data *Fisher-Pearson skewness*** $\equiv FPS_g = \dfrac{n^{3/2}(n-1)^{1/2}}{(n-1)(n-2)} \cdot g_g \sim= FPS_x$ (3c)

**Sample *Fisher-Pearson skewness*** $\equiv FPS_x \equiv \dfrac{n}{(n-1)(n-2)} \sum_{i=1}^{n}[(x_i - \bar{x})/s]^3$, *etc.* (3d)

$$FPS_x = \dfrac{n^{3/2}(n-1)^{1/2}}{(n-1)(n-2)} \cdot g_x > g_x \quad (3e)$$

Skewness constraints (3a), $g_g = g_x$, and (3c), $FPS_g = FPS_x$ *differ from the mean and variance constraints* (1b), (2b), *in that grouped data gamma skewness, $g_g$, and $FPS_g$ depend implicitly on the bin parameters $t_o$, h only through the shape*, $v_k$, *not on $t_o$ or h separately from shape*. (In (3a), $\bar{k}$ depends on $v_k$, i.e. $v_k(t_o, h; x_i)$.) Histogram *skewness level sets* for $g_g$, $FPS_g$ are *identical* to histogram *shape level sets*. Unlike histogram mean and variance, setting histogram skewness equal to data skewness does not lead to a line that may or may not intersect the shape level set parameter space. Histogram skewness does not vary continuously over a range so attempting to solve for ($t_o$, h) or solve for a shape, $v_k$, that leads to grouped data skewness, $g_g$ or $FPS_g$ that is the same as the data skewness, $g_x$ or $FPS_x$, almost never succeeds (except for symmetric histogram shapes for (a) *exactly* symmetric data or (b) other data, if any, with sample skewness equal to zero.) All of the histogram bin parameter values, ($t_o$, h), in the shape level set for $v_k$ lead to the same histogram skewness value, $g_g$ or $FPS_g$ and almost always this will *not* equal the data skewness, $g_x$ or $FPS_x$. Frequency and density histogram skewness is "shape skewness" and probably should be referred to as such. (This understanding of histogram skewness seems to be missing, incomplete or incorrect in Weber (2008**a**).)

(3d), (3e) above show a monotone relationship between sample gamma, $g_x$, and sample Fisher-Pearson coefficient, $FPS_x$. Consequently *rankings according to closeness of shape skewness, $g_g$ or $FPS_g$, to data skewness, $g_x$ or $FPS_x$, are the same for both sample gamma and Fisher-Pearson coefficient.* On account of a monotone relationship between histogram grouped data variance and histogram density variance, there is a monotone relationship between grouped data and density skewness so that shape skewness rankings vs. sample skewness also are the same. Thus all four combinations (histogram grouped data, density; *g*, *FPS*) identify the same *skewness-good* histogram shapes.[1, p 11]

**2.3 MOM UBW Frequency histogram example**. This and sections §2.4, §2.5 should break any lingering expectation of a unique MOM histogram. Consider MOM analysis via (1b), (2b), (3b) and shape level sets applied to illustrative data*, *n* = 12, {0.37, 1.13, 1.23, 2.25, 2.35, 2.45, 3.37, 4.37, 4.47, 5.37, 5.47, 5.61} (*originally created to have UBW histogram shapes that include (3,2,1,1,2,3) and (1,2,3,3,2,1), Weber 2008**a** Data #3). Calculating MOM estimates of histogram bin parameters ($t_o$, $h$) via mean, variance constraints, (1b), (2b), §2.2, is best understood from intersections of the straight lines from (1b), (2b) with shape level sets, Fig. A, Appen. A, Figs B.A-B.E, Appen. B. Histogram shape individual moment consistency means that straight lines (1b) or (2b) intersect the shape level set for the shape bin frequencies, $v_k$, used in (1b) and (2b). The Weber (2008a) data #3 has 123 uniform bin width histogram shapes of at most six bins. **A – E,** below, lists numbers of shapes for six kinds of mean and variance consistency and how this happens via moment constraint line intersections.

- **A. 44** shapes are *not* consistent with either the data mean or variance. Both mean and variance constraint lines, (1b), (2b) do *not* intersect the shape level set.
- **B. 17** shapes are consistent with mean but *not* variance; Mean constraint line intersects the shape level set, but variance line does not.
- **C. 11** shapes are consistent with variance but *not* mean; Variance constraint line intersects the shape level set, but mean line does not.
- **D. 32** shapes are *individually* mean and variance consistent, but *not* jointly consistent for the same bins. Mean and variance constraint lines intersect shape level set *but intersect each other outside the shape level set*. (Especially for **D**, but also for **A**, **B**, **C** intersection of (1b) and (2b) is outside of shape level set. So bins associated with the ($t_o$, $h$) intersection values do not lead to the same shape.)
- **E. 19** shapes are *jointly* mean and variance consistent, *for unique* ($t_o$, $h$) *and bins*; Mean and variance constraint lines intersect each other *inside the shape level set*.
- **F.** Shapes *very rarely* are consistent with data skewness. Shape skewness graphs are not straight lines like mean and variance. In (3a), skewness depends on ($t_o$, $h$) only through the shape, $v_k$. ($\bar{k}$ depends on $v_k$.) Skewness level sets are identical to shape level sets.

To further clarify, mean and variance constraints, (1b), (2b) lead to straight line *level curve* graphs. **(Even though histogram variance is *not linear in h*, it is independent of $t_o$. Variance constraint graphs are horizontal straight lines in {($t_o$, $h$)}.)** Outside of their associated shape level sets, ($t_o$, $h$) values that satisfy (1b), (2b) do not lead to associated shapes. However mean and variance constraint graphs outside of shape level sets *clarify* situations **A – E** by illustrating MOM estimates outside of a shape parameter space. Appendix **B** explains elementary calculations to identify **A – E**, including *line segment* solutions for **B**, **C** and **D**. §2.4, Tables 1ab, show the shapes for **B – E** and some of **F**.

**2.4 Venn Diagram for mean and variance consistency.** To further clarify mean and variance feasibilities, consider Fig. 1, a Venn Diagram of MOM Histogram Shapes for situations **A – E**.

A. $Sh/(M_g \cup V_g)$ – neither mean nor variance consistent
B. $M_g/V_g$ – mean but not variance consistent
C. $V_g/M_g$ – variance but not mean consistent
D. $(M_g \cap V_g)/J_g$ – mean, variance individually but not jointly consistent
E. $J_g$ – jointly mean and variance consistent

*$Sh$* = {*all* 123 shapes of at most six uniform width bins for Weber **2008a** data #3}
$M_g$ = {grouped data **m**ean consistent shapes}
$V_g$ = {grouped data **v**ariance consistent shapes}
$M_g \cup V_g$ = {grouped data mean or grouped data variance consistent shapes}
$Sh/(M_g \cup V_g)$ – neither mean nor variance consistent (A)
$M_g/V_g$ = {grouped data mean but not variance consistent shapes} (B)
$V_g/M_g$ = {grouped data variance but not mean consistent shapes} (C)
$(M_g \cap V_g)/J_g$ = {grouped data mean, variance *individually but not jointly* consistent shapes} (D)
$J_g$ = {grouped data *jointly mean and variance consistent* shapes} (E)

# Figure 1 - Venn Diagram of MOM Histogram Shapes

**Frequency Histogram Shape Moment Consistency Subsets**

$Sh = \{\,123\ Shapes\,\} = (M_g \cup V_g) \cup (Sh/(M_g \cup V_g))$

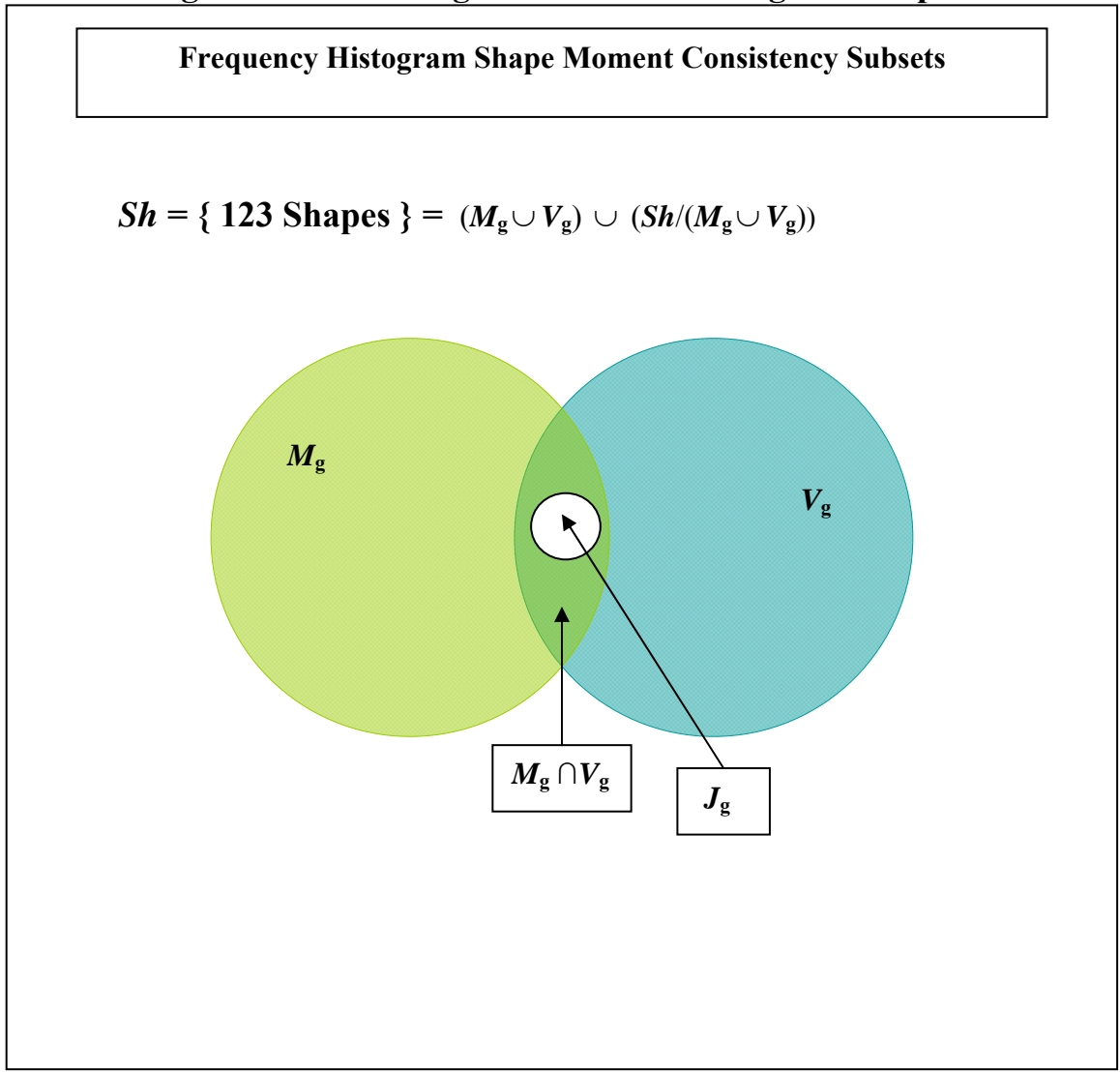

**Mean and Variance MOM Consistency Decomposition of Shapes**
For all of the shapes of at most six bins for Weber **2008a** Data #3

$Sh = J_g \cup ((M_g \cap V_g)/J_g) \cup (V_g/M_g) \cup (M_g/V_g) \cup (Sh/(M_g \cup V_g))$
$\#Sh = \#J_g + \#((M_g \cap V_g)/J_g) + \#(V_g/M_g) + \#(M_g/V_g) + \#(Sh/(M_g \cup V_g))$
$123 = 19 + 32 + 11 + 17 + 44$

$\#J_g = 19$ – jointly mean and variance consistent (E)
$\#((M_g \cap V_g)/J_g) = 32$ – mean, variance individually but not jointly consistent (D)
$\#(V_g/M_g) = 11$ – variance but not mean consistent (C)
$\#(M_g/V_g) = 17$ – mean but not variance consistent (B)
$\#(Sh/(M_g \cup V_g)) = 44$ – neither mean nor variance consistent (A)

---

$\#(M_g \cup V_g) = 79$ – mean or variance (or both) consistent  (123 – 44)

(To more explicitly compare MOM histogram and MOM Normal parameter estimation ia Venn Diagrams, see Appendix E.)

**2.5 Specific shape consistencies for 79 out of 123 shapes.** Out of 123 shapes, Table 1 segregates into eight columns 79 shapes that are consistent with the data mean or variance, with column headings as in Fig. 1: "$M_g$," "$V_g$," "$M_g \cap V_g$," "$J_g$," "Shape Skewness,", "$T_\gamma$," "$F_\gamma$," and "$J_g \cap T_\gamma$." The shapes in $M_g \cup V_g$ are not shown in a column. (A shape is in $M_g \cup V_g$ if it is in the column for $M_g$ *or* column for $V_g$ *or* both.) *The variety of shapes in column $J_g$ shows that jointly mean and variance consistent shapes are not similar.* (Joint consistency implies individual mean and variance consistency, but *not conversely*.)

$T_\gamma$ shows ranks of 10% of the $M_g \cup V_g$ shapes that are closest in gamma skewness to the data gamma skewness. Similarly $F_\gamma$ represents 5% of the shapes. Representing data as closely as possible in terms of matching graphic moments with data moments leads to the rightmost column, $J_g \cap T_\gamma$.

Ranking the shapes based on the deviation of shape skewness from data skewness identifies a skewness optimal shape. (Unfortunately the shapes that are closest in skewness, ranked –2, –1, 0, 1, 2, 3, 4, are neither mean nor variance consistent.)

**Table 1a**    $M_g \cup V_g \supseteq$    $M_g \cap V_g$      $\supseteq J_g$         $T_\gamma \supseteq$    $F_\gamma$      $J_g \cap T_\gamma$

**Method of Moments applied to Weber, J. S. (2008a) Data #3.**
**79 Mean or Variance consistent shapes, out of 123.**

| Page 1 | | | Data | Skewness ≡ $g_x$ = | -0.0288 | Skewness Rank of | Skewness Rank of | |
|---|---|---|---|---|---|---|---|---|
| | | | | $J_g$: Mean & | Shape | Shapes | Shapes | |
| $M_g$: Mean | $V_g$: Variance | $M_g \cap V_g$ | Variance **Jointly** | Skew | Within* | Within** | |
| Consistent | Consistent | Shapes | **Consistent** | ness | $T_\gamma$: Ten% | $F_\gamma$: Five% | |
| Shapes | Shapes | | | | of $g_x$ | of $g_x$ | |
| **12** | **12** | **12** | **12 – exact MISE** | **0** | | | |
| 1,11 | 1,11 | 1,11 | 1,11 | | | | |
| 2,10 | 2,10 | 2,10 | | | | | |
| 3,9 | 3,9 | 3,9 | 3,9 | | | | |
| 4,8 | 4,8 | 4,8 | | | | | |
| 5,7 | 5,7 | 5,7 | | | | | |
| 6,6 | 6,6 | 6,6 | **6,6** Rice,ShmMISE | 0 | **5** | **5** | **6,6** |
| 7,5 | 7,5 | 7,5 | 7,5 | | | | |
| 8,4 | | | | | | | |
| 9,3 | | | | | | | |
| 10,2 | | | | | | | |
| 11,1 | 11,1 | 11,1 | | | | | |
| 1,6,5 | 1,6,5 | 1,6,5 | 1,6,5 | | | | |
| 1,8,3 | | | | | | | |
| 1,10,11 | | | | | | | |
| 2,5,5 | 2,5,5 | 2,5,5 | | | | | |
| 2,7,3 | | | | -0.075 | **-8** | | |
| 3,4,5 | 3,4,5 | 3,4,5 | 3,4,5 | | | | |
| 3,5,4 | 3,5,4 | 3,5,4 | | | | | |
| 3,6,3 | 3,6,3 | 3,6,3 | | 0 | **5** | 5 | |
| 3,7,2 | | | | | | | |
| 3,8,1 | | | | -0.5482 | **-4** | -4 | |
| | 4,3,5 | | | | | | |
| | 4,4,4 | | | 0 | **5** | 5 | |
| 4,5,3 | 4,5,3 | 4,5,3 | | | | | |
| 5,3,4 | 5,3,4 | 5,3,4 | 5,3,4 | | | | |
| 5,4,3 | 5,4,3 | 5,4,3 | | | | | |
| 5,5,2 | | | | | | | |
| 6,3,3 | 6,3,3 | 6,3,3 | | | | | |
| 6,4,2 | | | | | | | |
| 6,5,1 | 6,5,1 | 6,5,1 | | | | | |
| 1,2,4,5 | 1,2,4,5 | 1,2,4,5 | | | | | |
| 1,3,3,5 | 1,3,3,5 | 1,3,3,5 | 1,3,3,5 | | | | |
| 1,4,2,5 | 1,4,2,5 | 1,4,2,5 | | | | | |
| 1,5,1,5 | 1,5,1,5 | 1,5,1,5 | **1,5,1,5** | -0.0762 | **-9** | | **1,5,1,5** |
| 1,5,2,4 | 1,5,2,4 | 1,5,2,4 | | | | | |
| 1,5,3,3 | 1,5,3,3 | 1,5,3,3 | 1,5,3,3 | | | | |
| 1,5,4,2 | | | | | | | |
| 1,5,5,1 | 1,5,5,1 | 1,5,5,1 | **1,5,5,1** | 0 | **5** | 5 | **1,5,5,1** |
| | 2,4,1,5 | | | | | | |
| | 3,4,2,4 | | | | | | |
| 2,4,3,3 | 2,4,3,3 | 2,4,3,3 | | | Almost | | |

**Table 1b** $M_g \cup V_g \supseteq$    $M_g \cap V_g$    $\supseteq J_g$      $T_\gamma \supseteq$   $F_\gamma$      $J \cap T_\gamma$

**Method of Moments applied to Weber, J. S. (2008a) Data #3.**
**79 Mean or Variance consistent shapes, out of 123.**

| Page 2 | | | Data | Skewness ≡ $g_x$ = | -0.0288 | Skewness **Rank** of Shapes Within* | Skewness **Rank** of Shapes Within** | |
|---|---|---|---|---|---|---|---|---|
| $M_g$: Mean Consistent Shapes | $V_g$: Variance Consistent Shapes | $M_g \cap V_g$ Shapes | $J_g$: Mean & Variance **Jointly Consistent** | Shape Skew Ness | | $T_\gamma$: Ten% of $g_x$ | $F_\gamma$: Five% of $g_x$ | |
| 3,3,1,5 | 3,3,1,5 | 3,3,1,5 | | | | | | |
| 3,3,2,4 | 3,3,2,4 | 3,3,2,4 | | -0.0491 | | -3 | | |
| 3,3,3,3 | 3,3,3,3 | 3,3,3,3 | **3,3,3,3** | 0 | | **5** | 5 | **3,3,3,3** |
| 3,4,2,3 | 3,4,2,3 | 3,4,2,3 | | | | | | |
| 3,4,3,2 | 3,4,3,2 | 3,4,3,2 | | | | | | |
| 3,4,4,1 | 3,4,4,1 | 3,4,4,1 | 3,4,4,1 | | | | | |
| 1,2,3,1,5 | 1,2,3,1,5 | 1,2,3,1,5 | 1,2,3,1,5 | | | | | |
| 1,2,3,2,4 | 1,2,3,2,4 | 1,2,3,2,4 | | | | | | |
| 1,2,3,3,3 | 1,2,3,3,3 | 1,2,3,3,3 | 1,2,3,3,3 | | | | | |
| 1,2,4,2,3 | 1,2,4,2,3 | 1,2,4,2,3 | 1,2,4,2,3 | | | | | |
| 1,2,4,4,1 | 1,2,4,4,1 | 1,2,4,4,1 | | | | | | |
| 1,3,3,2,3 | 1,3,3,2,3 | 1,3,3,2,3 | | | | | | |
| 1,3,3,4,1 | 1,3,3,4,1 | 1,3,3,4,1 | | | | | | |
| 1,4,2,2,3 | 1,4,2,2,3 | 1,4,2,2,3 | | | | | | |
| 1,4,2,4,1 | 1,4,2,4,1 | 1,4,2,4,1 | **1,4,2,4,1** | 0 | | **5** | 5 | **1,4,2,4,1** |
| 1,5,1,3,2 | | | | | | | | |
| 1,5,1,4,1 | 1,5,1,4,1 | 1,5,1,4,1 | 1,5,1,4,1 | | | | | |
| 2,3,2,2,3 | 2,3,2,2,3 | 2,3,2,2,3 | | | | | | |
| 2,4,1,2,3 | | | | | | | | |
| 2,4,1,3,2 | | | | | | | | |
| | 3,2,2,2,3 | | | 0 | | **5** | 5 | |
| 3,3,1,2,3 | 3,3,1,2,3 | 3,3,1,2,3 | | | | | | |
| 3,3,1,3,2 | | | | | | | | |
| 3,3,1,4,1 | 3,3,1,4,1 | 3,3,1,4,1 | | | | | | |
| 1,2,3,1,2,3 | 1,2,3,1,2,3 | 1,2,3,1,2,3 | **1,2,3,1,2,3** | -0.0552 | | **-6** | | **1,2,3,1,2,3** |
| 1,2,3,1,3,2 | 1,2,3,1,3,2 | 1,2,3,1,3,2 | | -0.0859 | | **-11** | | |
| 1,2,3,1,4,1 | 1,2,3,1,4,1 | 1,2,3,1,4,1 | | | | | | |
| | 1,2,3,2,3,1 | | | | | | | |
| 2,1,3,1,2,3 | 2,1,3,1,2,3 | 2,1,3,1,2,3 | | | | | | |
| | 2,1,3,2,2,2 | | | | | | | |
| | 2,2,2,2,3,1 | | | | | | | |
| 2,2,3,1,3,1 | 2,2,3,1,3,1 | 2,2,3,1,3,1 | | | | | | |
| **3,0,3,1,2,3** | **3,0,3,1,2,3** | **3,0,3,1,2,3** | ( Exact <u>ML</u> ) | | | | | |
| 3,1,3,1,2,2 | | | | | | | | |
| | 3,1,3,2,2,1 | | | | | | | |
| | 3,2,2,2,2,1 | | | | | | | |
| | 3,3,1,2,2,1 | | | | | | | |

\* Within 5% means +/- 5% of 123 shapes unless max/min = zero or a tie, etc.
That is, six shapes less than and six greater than the data gamma statistic

\* Within 10% means +/- 10% of 123 shapes unless max/min = zero or a tie, etc.
That is, thirteen shapes less than and thirteen greater than the data gamma statistic

## 2.6 Agreement of ALL frequency histogram grouped data moments.

*For completeness*, note that arithmetic progressions in the data lead to countable sets of bins, $h > 0$, for which *all frequency* histogram grouped data moments agree with *all* of the data moments. Real world measurements are represented by rational numbers: $x_i = p_i/q_i$, for relatively prime integers $p_i$, $q_i$. Let $Q \equiv$ least common multiple of the integers $q_i$, "LCM($q_i$)," so that $x_i = k_i p_i/Q$, $k_i = Q/q_i$. For bin widths $h^z \equiv 1/zQ$, $z = 1, 2, 3, \ldots \infty$, and $t_o^z \equiv x_{min} - h^z/2$, the bin mid points are identical to data values with frequencies equal to data value frequencies, so *all* frequency histogram grouped data moments are *based on* the *same* set of *values* and value frequencies *as the data*. This augments the well known* fact that as $h \rightarrow 0$, histogram moments converge to data moments. (*-email from David Scott). Also gamma and Fisher – Pearson skewness statistics will agree. In rare situations wherein $Q$ may be relatively small, there may be interesting examples. (See also Weber, J. S. et al (R. Stong) 2005/2006)

Since histogram density variance is $h^2/12$ greater than frequency histogram variance, from the perspective of arithmetic progressions in the data, histogram density variance converges to grouped data variance as $h \rightarrow 0$ for $h^z$, as $z \rightarrow \infty$, etc.

[1] [p 6] This is clear in (3a) - (3d). Gamma skewness and Fisher-Pearson coefficient for data, $x_i$, and frequency and density histograms are invariant for affine transformations of the data and histogram bin midpoints and endpoints. Hence histogram skewness is invariant over a shape level set: $(t_o, h) \rightarrow (a^t t_o + b^t, a^h h + b^h)$, $a^t, a^h > 0$, $(a^t t_o + b^t, a^h h + b^h)$ ε SLS. Since $a^t, b^t, a^h, b^h$ cancel/drop out, $\gamma_x(a^t t_o + b^t, a^h h + b^h) = \gamma_x(a^t t_o + b^t, a^h h + b^h)$ just as $\gamma_x(ax_i + b) = \gamma_x(x_i)$, $\gamma_g(v_k(a^t t_o + b^t, + (k - ½)(a^h h + b^h))) = \gamma_g(v_k(t_o + (k - ½)h)$; just as $FPS_x(ax_i + b) = FPS_x(x_i)$, $FPS_g(v_k(a^t t_o + b^t, + (k - ½)(a^h h + b^h))) = FPS_g(v_k(t_o + (k - ½)h)$. <u>Changing the bin parameters *within a level set* leads to an affine transformation of the bin midpoints</u>. That is $(t_o, h) \rightarrow (t^\#_o, h^\#)$ and is $v_k$ unchanged, but bin midpoints change from $(t_o + (k - ½)h)$ to $(t_o + (t^\#_o - t_o) + [(k - ½)(h^\#/h)]h) = (t^\#_o + (k - ½)h^\#)$, so $a = (h^\#/h)$ and $b = (t^\#_o - t_o)$. Skewness is dependent only on the shape bin frequencies, $v_k$, and stays the same. <u>In EXCEL® data, $x_i$, and $ax_i + b$, $a > 0$, e.g. $x_i$ and $27x_i + 41$, lead to the same EXCEL skewness descriptive statistics</u>.

NOTE regarding references: All of this began with the simple question: "What uniform bin width histogram shapes can data have?" The large number of references reflects a diligent search for any thinking on this. It is doubtful there is anything in print on this question or shape level sets. Few of the references below need to be included in a revision.

NOTE regarding references: All of this began with the simple question: "What uniform bin width histogram shapes can data have?" The large number of references reflects a diligent search for any thinking on this. It is doubtful there is anything in print on this question or shape level sets. Few of the references below need to be included in a revision.

**Appendices**.
A. Shape level sets – Weber, J. S. (2016) "Histogram bin edge discontinuity." In review process.
B. Computational details to identify E; A, B, C, D; line segments of ($t_o$, $h$) values for B, C, D.
C. Pseudo code/ list of steps: Shape level set and MOM histogram analysis.
D. Summary of mean, variance and skewness formulae
E. Fig. E Venn Diagram for MOM Normal parameter estimation.

**Appendix A. Uniform Bin Width Histogram Shape level sets**

Weber, J. S. (2008**a**), Weber, J. S. (2016) "Histogram bin edge discontinuity" in review process. Histogram shape is a list of bin counts, ($v_k$), or relative frequencies, ($v_k/n$), or density *step function* values, ($v_k/nh$), $k = 1$ to **K**. For fixed data, $x_i$, shape level sets (SLS) are bin location and width values, ($t_o$, $h$), that lead to half-open bins $\{[t_o + (k-1)h, t_o + kh) \mid k = 1 \text{ to } K\}$ that lead to the same uniform bin width histogram shape. ("Shape level sets" could be called "Shape inverse-images in $\{(t_o, h)\}$" or "Shape pre-images," *etc.*)

SLSs for *three or more bins* are interiors of convex polygons with some but not all boundary points, i.e. some but not all vertices and edges: bounded, but not compact, neither open nor closed. Edges, vertices belong to a SLS *or* an adjacent SLS, etc. (SLS for 1 or 2 bin shapes are unbounded half-plane intersections.)

SLSs were considered first to answer the question "What UBW histogram shapes are possible for data, $x_i$?" Beyond answering that question, SLSs are essential in exactly calculating many kinds of optimal or other kinds of histograms. More importantly, exact calculations *unexpectedly* show that *reasonable compelling approximations* are too inaccurate, challenging some thoughts about histograms (Weber, J. S., 2016, In review. "Histogram bin edge discontinuity")

Having *all* shapes is key in most uses of SLSs so construction of SLSs must *clearly, explicitly* lead to *the exhaustive set of UBW histogram shapes* for any maximum number of bins, **K**.

Discussions of histograms often arbitrarily index UBW bins, independently of the data. Consequently shape is associated with *many* doubly infinite sequences of non negative integer bin counts, $v_k$, $-\infty < k < \infty$. (E.g. Scott, D. W. 1992). Associating different sequences (e.g. $v^*_k = v_{k+1}$, $v^{**}_k = v_{k+2}$, $-\infty < k < \infty$ etc.) with essentially one shape leads to translations and transformations of shape level set cells in $\{(t_o, h)\}$. Shape should not be ambiguous. Shape level sets should be only one cell. Obtaining shape level sets that are one cell instead of many is done via a set $D_o \subset \{(t_o, h)\}$ by defining shape as the unique list of bin frequencies, $v_k$, $k = 1$ to $K$, with $K$ large enough so that $\sum_{k=1 \text{ to } K} v_k = n =$ the number of data points, (**A1**):

$$v_k, k = 1 \text{ to } K, 0 \leq v_k, 1 \leq v_1, \sum_{k=1 \text{ to } K} v_k = n \text{ (and } k < 1 \text{ or } K < k \Leftrightarrow v_k = 0\text{)} \quad (\textbf{A1})$$
$$\underline{\textit{\textbf{not}}} \; v_k, -\infty < k < +\infty, \; 0 \leq v_k, \; \sum_{-\infty < k < \infty} v_k = n$$

Defining shape as the unique list $v_k$, $k = 1$ to $K$, etc., leads to $D_o$ in $\{(t_o, h)\}$, via constraints (**A2**), (**A3**). The first bin contains the minimum data value, $x_{\min}$, (**A2**), and a bin indexed *at most*\* **K** contains $x_{\max}$, (**A3**) (\**or exactly* the $K^{\text{th}}$ bin, similar to (**A2**).)

$$t_o \leq x_{\min} < t_o + h \quad (\textbf{A2})$$
$$x_{\max} < t_o + Kh \quad (\textbf{A3})$$

It is helpful and not misleading to impose a *large* bound on bin width:
$$0 < h \leq (x_{max} - x_{min}) + \delta, \ 0 < \delta \tag{A4}$$
(Arbitrarily bounding bin width, $h$, by less than the range of the data, $x_i$, for example, $h \leq$ (data range)/2 $\equiv (x_{max} - x_{min})/2$, etc, sometimes excludes *optimal* UBW bin parameter values, for example MISE histograms, §2.4, Table 1a, top row. (Also Weber, J. S. (2016) "Histogram bin edge discontinuity", in review.)

Constraints (**A2**) – (**A4**) define $D_o$ as a bounded subset of $\{(t_o, h)\}$ that contains *all* of the points, $(t_o, h)$ leading to UBW histogram *shapes with three or more bins*, $K \geq 3$, and *subsets* of the points $(t_o, h)$ leading to shapes with one or two bins, $K = 1$ or 2. (Eliminating (**A4**) leads to unbounded $h$, unbounded $D_o$ and *all* $(t_o, h)$ points for all SLSs.)

The overall procedure is:

    **1**. Calculate $D_o$ vertices from (**A2**), (**A3**), (**A4**).

    **2**. Calculate shape level set vertices from shape level set boundaries, (**A5**), below. Shape level set vertices arise from intersections of lines (**A5**) with each other and with the boundaries of $D_o$.
$$t_o + kh = x_i, \ k = 1 \text{ to } K, \ i = 1 \text{ to } n \tag{A5}$$
(or $i^* = 1$ to $n^*$ for $n^*$ distinct values, $x^*_{i*}$)

In words, (**A5**) means (**A6**)
$$bin\ edge = data\ value \tag{A6}$$

Lines (**A5**) *partition* $D_o$ into shape level set cells.

    **3**. Calculate UBW histogram shapes.

    **4**. Calculate, examine MOM histograms, MISE histograms, etc.

Figure A illustrates a shape level set in $D_o$. ($D_o$ is not shown.)

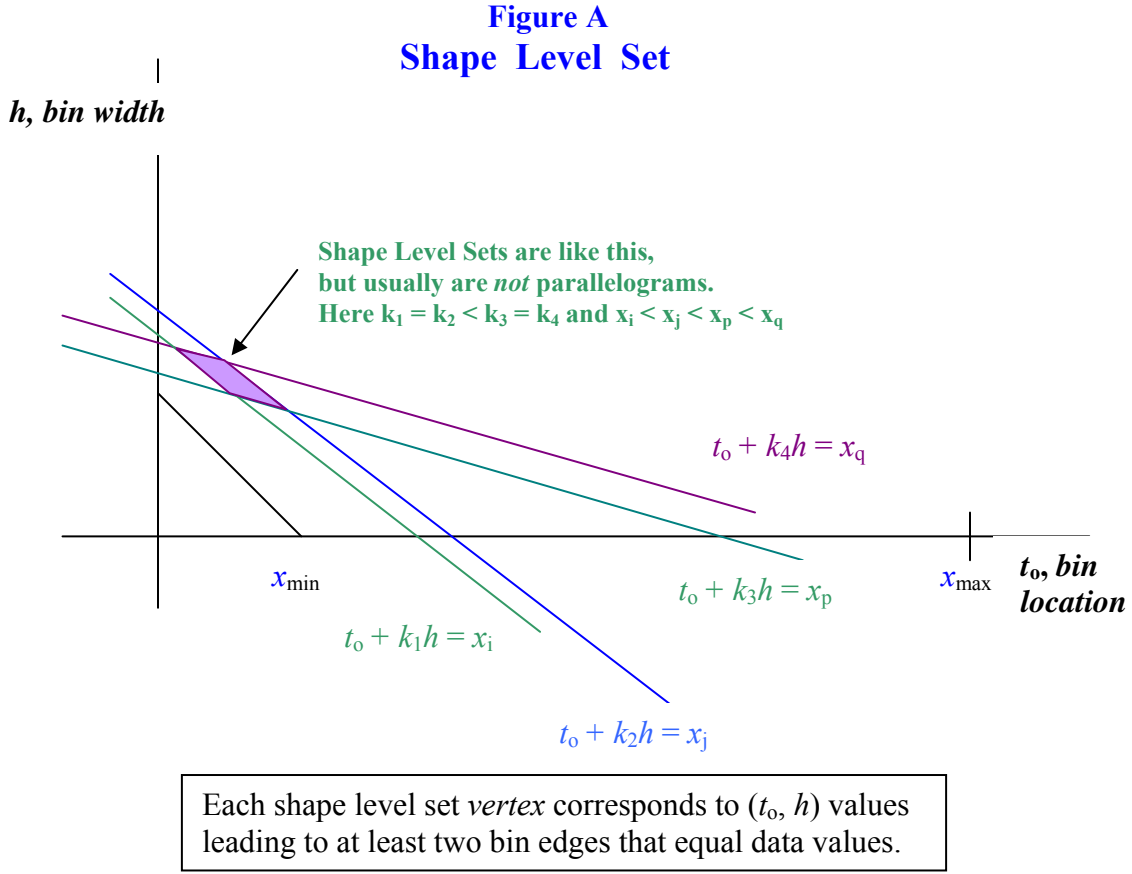

**Figure A**
**Shape Level Set**

Shape Level Sets are like this, but usually are *not* parallelograms. Here $k_1 = k_2 < k_3 = k_4$ and $x_i < x_j < x_p < x_q$

Each shape level set *vertex* corresponds to ($t_o$, $h$) values leading to at least two bin edges that equal data values.

*To insure computational reproducibility* by avoiding ambiguous classification of data that equal bin edges (i.e., , $x_i = t_o + kh$), shapes are determined from SLS vertices via a convenient interior point, such as the average of the first three vertices, (**A7**), average of all of the vertices, (**A8**), or any strictly convex combination of three or more vertices or two nonadjacent vertices.

$$t^{s, \text{int}}_o = (t^{s, 1}_o + t^{s, 2}_o + t^{s, 3}_o)/3, \ h^{s, \text{int}} = (h^{s, 1} + h^{s, 2} + h^{s, 3})/3 \text{ for the } s^{\text{th}} \text{ SLS} \quad \textbf{(A7)}$$

or

$$t^{s, \text{int}}_o = (\sum_{j = 1 \text{ to } V_s} t^{s, j}_o)/V_s, \ h^{s, \text{int}} = (\sum_{j = 1 \text{ to } V_s} h^{s, j})/V_s, \quad \textbf{(A8)}$$
for $V_s$ vertices for the $s^{\text{th}}$ SLS

Once shapes have been determined using (**A7**) or (**A8**), how can the results be presented prior to determining various optimal or other kinds of histograms?

For the $s^{th}$ shape, $s = 1$ to $S$, concatenating number of bins, $K_s$, bin counts, $v_{s,k}$, and SLS vertices, $(t_o, h)_{s,v}$, leads to **(A9)**. Sorting **(A9)** lexicographically on $K_s$, then $v_{s,k}$ shapes gives a canonical presentation of shapes and shape level set vertices.

$$\{(K_s, v_{s,k}, (t_o, h)_{s,v}) \mid s = 1 \text{ to } S, k = 1 \text{ to } K_s, v = 1 \text{ to } V_s\} \quad \textbf{(A9)}$$

$S \equiv$ number of shapes, $K \equiv$ max number of bins
$K_s \equiv$ number of bins for $s^{th}$ shape ($\equiv$ index of bin for $x_{max} \leq K$)
$V_s \equiv$ number of vertices for $s^{th}$ shape

**(A9)** is a right ragged $S \times (1 + K_s + 2V_s)$ matrix, $S$ rows, $(1 + K_s + 2V_s)$ entries in each row.

Finally, before leaving the discussion of shape level sets, we revisit $D_o$ in more detail. As already noted, three conditions, **(A2), (A3), (A4)** define $D_o$ in $\{(t_o, h)\}$:

1 - **(A2)** - $x_{min}$ contained in a first bin: $t_o \leq x_{min} < t_o + h \leftrightarrow$
    1a: $t_o \leq x_{min}$
    1b: $x_{min} < t_o + h$

2 - **(A3)** - <u>At most $K$ bins</u> $\leftrightarrow$
    2a: $x_{max} < t_o + Kh$
    (Or <u>exactly $K$ bins</u> $\leftrightarrow x_{max}$ is in the last bin $[t_o + (K-1)h, t_o + Kh) \leftrightarrow$
    2b: $t_o + (K-1)h \leq x_{max}$ and 2a.)

3 - **(A4)** - $D_o$ is bounded: $h < B \equiv (x_{max} - x_{min}) + \delta, 0 < \delta$

These lead to the following *boundaries* for $D_o$:

    1a: $t_o = x_{min}$ for $t_o \leq x_{min}$
    1b: $x_{min} = t_o + h$ for $x_{min} < t_o + h$
    2a: $x_{max} = t_o + Kh$ for $x_{max} < t_o + Kh$
    3: $h = (x_{max} - x_{min}) + \delta$ for $h \leq (x_{max} - x_{min}) + \delta$

These boundaries lead to four vertices for $D_o$ (or "$D_o^{\leq K}$" or, with modification "$D_o^{=K}$"), clockwise:

***vertex*** $_1$: (1a,3): $(x_{min}, (x_{max} - x_{min}) + \delta)$
***vertex*** $_2$: (1a,2): $(x_{min}, (x_{max} - x_{min})/K)$
***vertex*** $_3$: (1b,2): $(x_{min} - (x_{max} - x_{min})/(K-1), (x_{max} - x_{min})/(K-1))$
***vertex*** $_4$: (1b,3): $(x_{min} - ((x_{max} - x_{min}) + \delta), (x_{max} - x_{min}) + \delta)$

<span style="color:green"><u>*Normalizing data*</u> **[ $x_{min}$, … …, $x_{max}$ ] *to* [0,1] and letting $K \to \infty$, $\delta \to 0$ brings *vertex* $_2$ and *vertex* $_3$ together, etc. leading to $D_o^{\infty}{}_{[0,1]}$:**</span>

***vertex*** $_1$: (1a,3): $(x_{min}, (x_{max} - x_{min}))$ ; $(0, 1 + \delta)$      $\to$ (0, 1)
***vertex*** $_2$: (1a,2): $(x_{min}, 0)$    ; $(0, 1/K)$      $\to$ (0, 0)
***vertex*** $_3$: (1b,2): $(x_{min}, 0)$    ; $(-1/(K-1), 1/(K-1))$      $\to$ (0, 0)
***vertex*** $_4$: (1b,3): $(x_{min} - ((x_{max} - x_{min}))), (x_{max} - x_{min}))$ ; $(-(1+\delta), 1+\delta)$      $\to$ (−1, 1)

So $D_o^{\infty}{}_{[0,1]}$ is the triangle with vertices: **(0,0), (0,1), (-1,1)**.

Although normalizing data to [0,1] may reveal notable features, obviously $x_{min}$, $x_{max}$ and $(x_{max} - x_{min})$ are temporarily out of view.

**Appendix B. Computational details for situations A through E leading to Table 1.**

After shape level sets and shapes are obtained, what is the next step for MOM? How are situations **A – E** identified?

    **1**. Test for **E**, joint mean and variance consistency, by solving (1b), (2b) for unique $t_o$, $h$ and verifying that $t_o$, $h$ lead to bins that reproduce the shape, $v_k$.

    **2**. If not **E**, then identify situations **A**, **B**, **C**, or **D** by sign changes in $m$, $v$ evaluations at the SLS vertices.

|  | $m(t_o, h)$ | |
|---|---|---|
| $v(t_o, h)$ | no sign change | sign change |
| no sign change | **A** - $Sh/(M_g \cup V_g)$ | **B** - $M_g / V_g$ |
| sign change | **C** - $V_g / M_g$ | **D** - $M_g \cap V_g / J_g$ |

Below are objective functions $m(t_o, h)$, $v(t_o, h)$, $sk(t_o, h)$ for mean, variance and skewness for frequency* histogram grouped data:

$$m(t_o, h) = \text{grouped data mean}(t_o, h; x_i) - \text{data mean} = t_0 + h(\bar{k} - \frac{1}{2}) - \bar{x}$$

$$v(t_o, h) = \text{grouped data variance}(t_o, h; x_i) - \text{data variance} = \frac{h^2}{n-1}[\sum_{k=1}^{K} v_k(k - \bar{k})^2] - s_x^2$$

$$sk(t_o, h) = \text{grouped data skewness}(t_o, h; x_i) - \text{data skewness} = \frac{h^2}{n-1}[\sum_{k=1}^{K} v_k(k - \bar{k})^2] - g_x$$

(*for histogram densities, histogram density variance = frequency histogram variance + $h^2/12$.) <u>Actual graphing is not done</u> for the level curves for $(t_o, h)$ points that have grouped data mean and variance equal to data mean and variance, and looking for level curve intersections with corresponding shape level sets. Either situation **E** is determined, or sign changes in $m$, $v$ signal identify situation **A**, **B**, **C** or **D**. Objective functions $m(t_o, h)$ and $v(t_o, h)$ are evaluated at the vertices of each shape level set. If there is a sign change, then there is level curve - level set intersection, otherwise there isn't. This gives the main idea. Of course there are extra details such as dealing with $m$, $v$ evaluations of zero at a vertex when there is no strict sign change, and, if wanted, calculating line segment solution sets for **B**, **C**, **D**. Since all of the vertices and objective function evaluations at vertices are rational numbers** for real world observations represented as rational numbers, exact arithmetic, *exact* tests of zero can be done, etc.)

This is important for renewed interest in robust computational reproducibility. (Stodden et al, 2013). (** except **E** – see (2b)).

When there is a sign change, a convex combination of vertices with different signs for $m(t_o, h)$ or $v(t_o, h)$ locates the zero value $(t_o, h)$ point for $m(t_o, h)$ or $v(t_o, h)$ on a line connecting vertices with different signs, towards obtaining line segments of $(t_o, h)$ values for **B**, **C**, **D**, if wanted.

Skewness objective function $sk(t_o, h)$, is similar to $m(t_o, h)$, $v(t_o, h)$, except is the same for all $(t_o, h)$ in a level set. The value of $sk(t_o, h)$ is the actual deviation, plus or minus, of a *shape skewness* with the data skewness. Shapes can be ranked by $sk(t_o, h)$. Histogram shapes with skewness same sign as $g_x$ that are close to the data skewness, i.e. relatively small $sk(t_o, h)$, can be considered *skewness-good*.

Table 1 in the main text above, shows jointly mean and variance consistent shapes scoring well on $sk(t_o, h)$ via rankings, in the right most column.

Ranked comparisons with data skewness are the same for $\gamma$ and FPS skewness, the same for frequency and density histograms and all lead to the same skewness ranking of UBW shapes when compared to data skewness

If there is still lack of clarity regarding situations A, B, C, D, E, Figure A has been revised to include mean ("M") and variance ("V") constraint lines, noted as:

Fig B.A – neither mean, M, nor variance, V, consistent

Fig B.B – mean but not variance consistent

Fig B.C – variance but not mean consistent

Fig B.D – mean and variance independently but not jointly consistent

Fig B.E – mean and variance jointly consistent

Note that mean constraint lines, (1b), "M"

$$\boldsymbol{h} = \bar{x}/(\bar{k} - \frac{1}{2}) - t_0/(\bar{k} - \frac{1}{2}); \quad x_{min} + h < t_0 \leq x_{min} \qquad (1b) \text{ - (}\boldsymbol{M}\text{)}$$

always have negative slope in the $\{(t_o, h)\}$ plane. The variance constraint lines, (2b), "V"

$$\boldsymbol{h} = [(n-1) s^2_x / \sum_{k=1}^{K} v_k (k - \bar{k})^2]^{\frac{1}{2}} + \boldsymbol{0}\, t_o \qquad (2b) \text{ - (}\boldsymbol{V}\text{)}$$

always are horizontal.

**Figure B.A: neither M nor V**

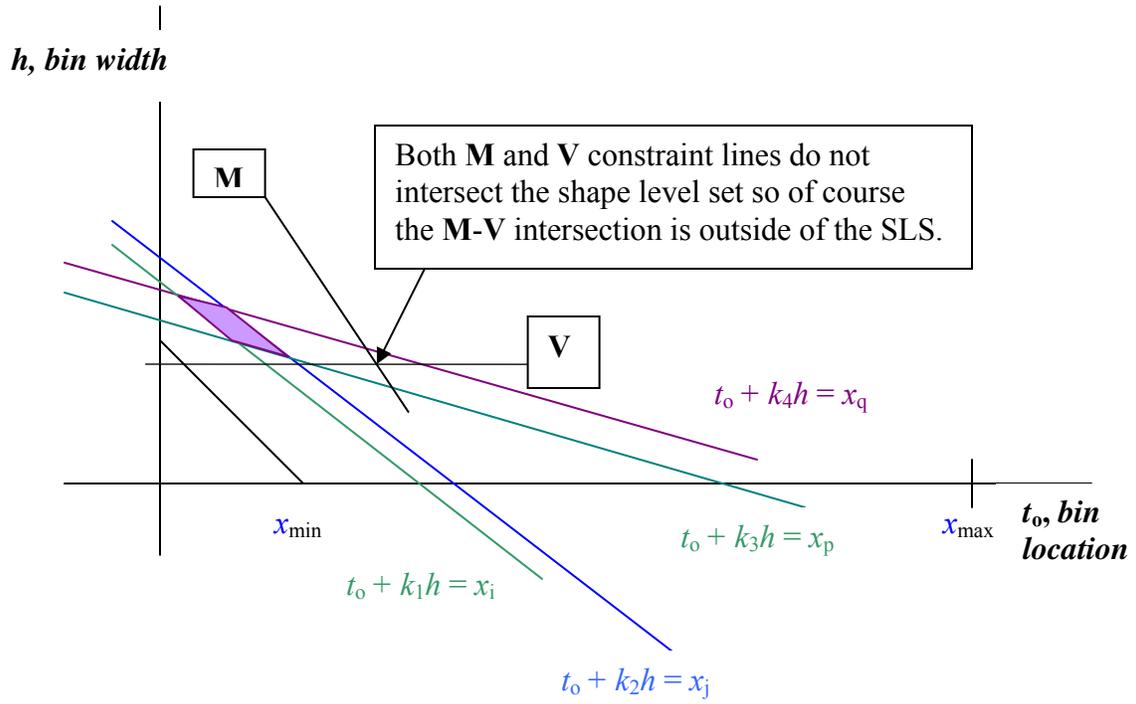

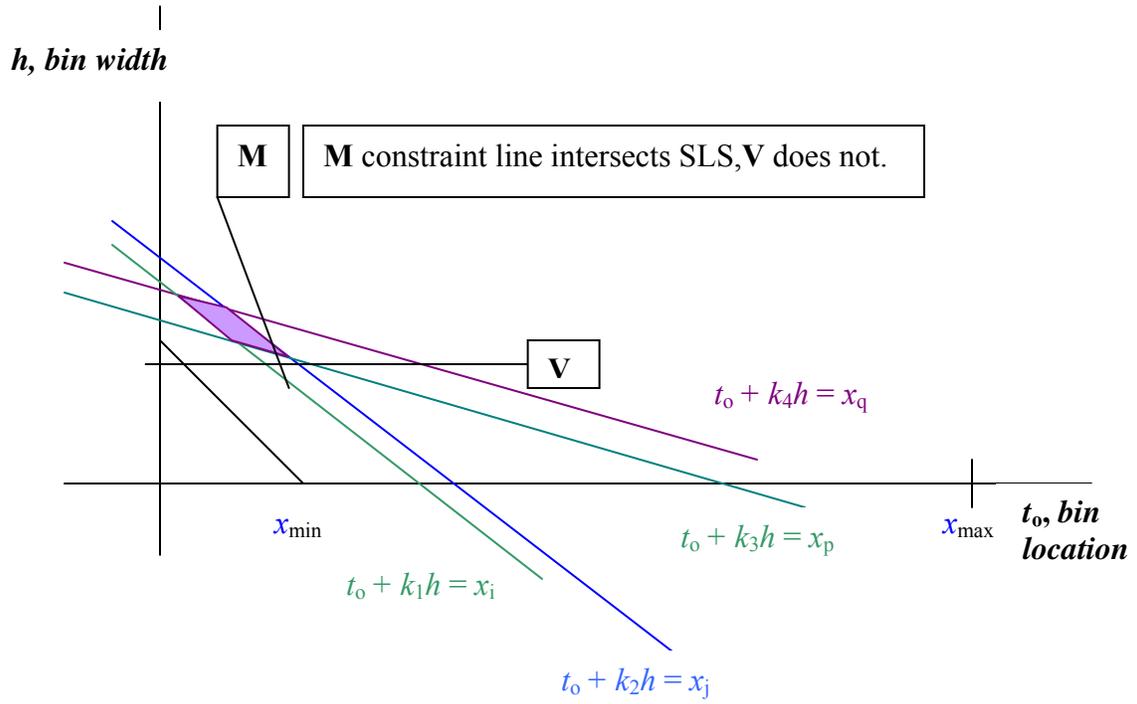

**Figure B.B:** M, not V

**Figure B.C: V, not M**

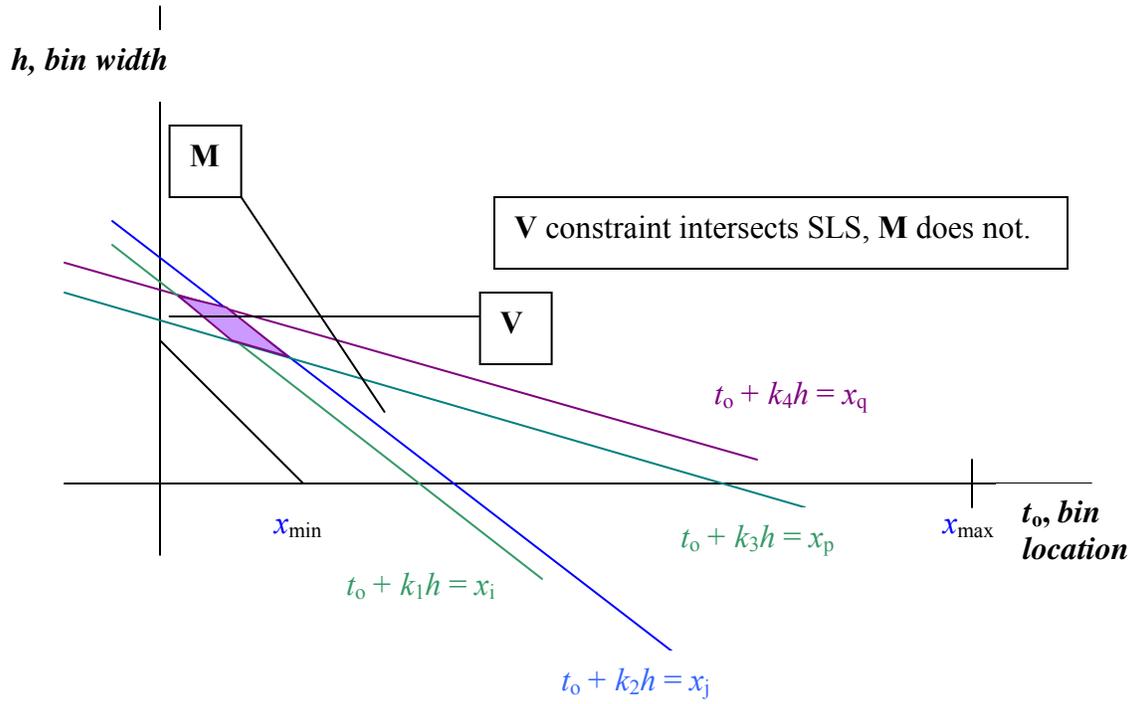

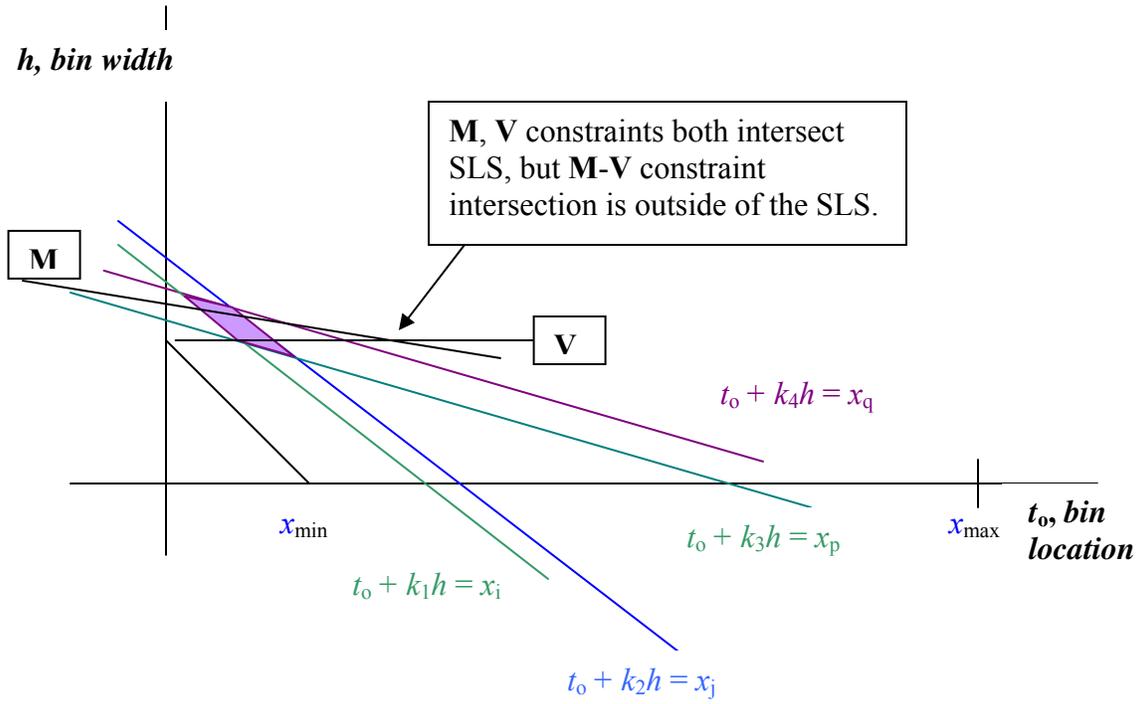

Figure B.D: M,V independently, not jointly

**Figure B.E: joint consistency**

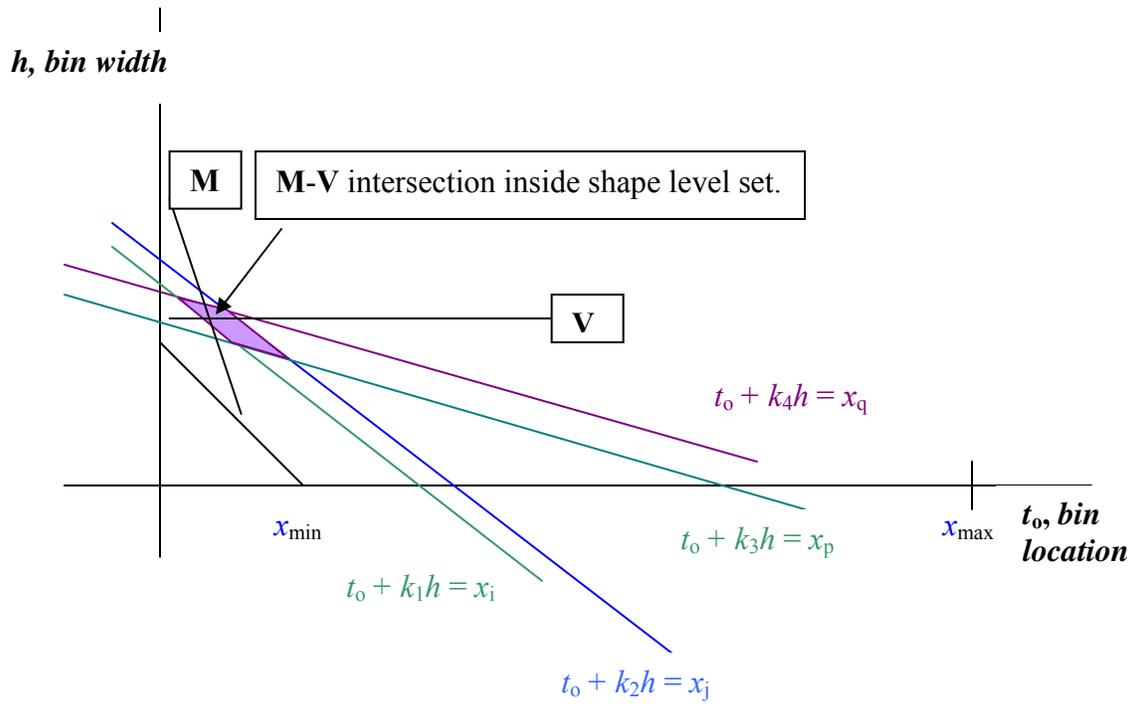

**APPENDIX C Shape level set MOM histogram analysis <u>Pseudo code/ list of steps</u>**.
  I. **User Input**
    A. Read data, $x_i$, $i = 1$ to $n$ = number of data points.
    B. Specify range of numbers of bins, if not program determined: 1 to $K(data)$.
    C. Specify desired views & reports.
  II. **Shape Level Sets Algorithm steps**
    A. Determine data value set, $x^*_{i^*}$, $i^* = 1$ to $n^*$ = number of distinct values.
    B. In $\{(t_o, h)\}$ determine $D_o$ vertices from $x_{min}$, $x_{max}$, $K$.
    C. In $\{(t_o, h)\}$ generate lines $t_o + kh = x^*_{i^*}$, $k = 1$ to $K$, $i^* = 1$ to $n^*$.
      For each line determine new vertices from intersection of line with $D_o$ and existing previously determined shape level sets leading to new shape level sets, $(t^{s,k}_o, h^{s,k}) \mid k = 1$ to $K_s$ = the number of vertices for the $s^{th}$ shape$\}$
      (When last line, $t_o + Kh = x^*_{n^*}$, $k = 1$, is processed., then level sets for all of the shapes of at most $K$ bins will be specified as sets of vertices.)
    D. Cycle through the set of shape level set vertices to determine the shape associated with each:
      1. determine a convenient SLS interior point, $(t^{s,int}_o, h^{s,int})$, such as the average of the vertices $\{(t^{s,v}_o, h^{s,v}) \mid v = 1$ to $V_s$, for each SLS, $s = 1$ to $S$.
      2. determine the bin counts, $v^s_k$, $k = 1$ to $K$. from bins:
        $[t^{s,int}_o + (k-1)h^{s,int}, t^{s,int}_o + kh^{s,int})$ and data, $x_i$, $i = 1$ to $n$.
      3. for each shape, $s = 1$ to $S$, determine $K_s \equiv$ max $k$ s.t. $v^s_k > 0$, $k = 1$ to $K$.
    E. Sort (**A9**) lexicographically, ascending on $K_s$, $v_{s,k}$.

$$\{(K_s, v_{s,k}, (t_o, h)_{s,v}) \mid s = 1 \text{ to } S, k = 1 \text{ to } K_s, v = 1 \text{ to } V_s\} \qquad (\mathbf{A9})$$

    $S \equiv$ number of shapes, $K \equiv$ max number of bins
    $K_s \equiv$ number of bins for $s^{th}$ shape ($\equiv$ index of bin for $x_{max} \leq K$)
    $V_s \equiv$ number of vertices for $s^{th}$ shape
  (**A9**) is a right ragged $S \times (1 + K_s + 2V_s)$ matrix, $S$ rows, $(1 + K_s + 2V_s)$ entries in each row.
  III. **MOM Analysis Algorithm steps**
    A. For bin counts for each shape, $v_{s,k}$, $s = 1$ to $S$, with (1b), (2b) to determine if $s^{th}$ shape is situation **E**. If not then evaluate $m(.)$, $v(.)$ at vertices $\{(t^{s,v}_o, h^{s,v}) \mid v = 1$ to $V_s\}$, to identify situations **A**, **B**, **C**, **D**. Determine straight line solution sets for **B**, **C**, **D**.
    B. Calculate *gamma* or *FPS* skewness for the data, $x_i$, and for all of the shapes. Rank the shapes according to deviation from the data skewness. Identify the 5% or 10% closest and greater than and 5% or 10% closest and less than data skewness and same sign. Rank these in absolute deviation, identify any that satisfy mean and variance constraints exactly (situation **E**), or situations **D**, **C**, **B**, etc, as desired.
    C. Create <u>Table 1</u> for further examination.
  IV. **Other kinds of histograms**. Use (**A9**) to determine other exact histograms such as MISE, maximum likelihood, etc., by evaluating various statistical objective functions for each shape, ranking the shapes according to exact objective function values to identify a global optimum, etc. (Weber, J. S. (2016) "Bin Edge Discontinuity" In review.)

**Actual operation** is **IAB**, **II**, **III**, **IV**, then **IC**, since **II**, **III**, **IV** calculate instantly and **IC** simply selects and displays calculated results for a GUI (graphical user interface).

APPENDIX D: Method of Moments Formulae, etc.

| Moments & Other Selectors | Sample | Grouped Data, "…g" | Histogram Density, "…Hd" | Notes |
|---|---|---|---|---|
| Mean | 1. $\bar{x} = \dfrac{1}{n}\sum_{i=1}^{n} x_i$ | 2. $\bar{x}_g = \dfrac{1}{n}\sum_{k=1}^{K} v_k [t_0 + (k - \tfrac{1}{2})h]$ $= t_0 + h(\bar{k} - \tfrac{1}{2})$ | 3. $\bar{x}_{Hd} = \dfrac{1}{n}\sum_{k=1}^{K} v_k [t_0 + (k - \tfrac{1}{2})h]$ $= t_0 + h(\bar{k} - \tfrac{1}{2})$ | 1. Histogram Density mean = Grouped data mean. 2. Depends on $t_0$, h through individual values of $t_0$, h, as well as the shape, i.e. bin counts. |
| Variance | 4. $s^2 = \dfrac{1}{(n-1)}\sum_{i=1}^{n}(x_i - \bar{x})^2$ | 5. $s^2_g = \dfrac{h^2}{n-1}[\sum_{k=1}^{K} v_k (k - \bar{k})^2]$ "SSR" | 6. $\sigma^2_{Hd} = \sigma^2_g + \dfrac{h^2}{12}$ "SSR" + "SSE" | 1. Histogram Density variance = grouped data second moment + bin uniform density variance. 2. Depends on $t_0$, h through h and the shape, i.e. bin counts, but not to $t_0$ by itself. |
| Skewness | 7. $g \equiv \dfrac{\tfrac{1}{n}\sum_{i=1}^{n}(x_i - \bar{x})^3}{(\tfrac{1}{n}\sum_{i=1}^{n}(x_i - \bar{x})^2)^{3/2}}$ | 8. $g_g \equiv \dfrac{\tfrac{1}{n}\sum_{k=1}^{K} v_k (k - \bar{k})^3}{(\tfrac{1}{n}\sum_{k=1}^{K} v_k (k - \bar{k})^2)^{3/2}}$ | 9. $g_{Hd} \equiv \dfrac{\tfrac{1}{n}\sum_{k=1}^{K} v_k (k - \bar{k})^3}{(\tfrac{1}{n}\sum_{k=1}^{K} v_k (k - \bar{k})^2 + \tfrac{1}{12})^{3/2}}$ | 1. **Almost** the same for Grouped data & Histogram Density. 2. Depends on $t_0$, h only through the shape, i.e. the bin counts. 3. Since based on a density standardized $3^{rd}$ moment, satisfies R.A.G.(1991), p388 P1, P2, P3, P4. |
| ML | NA | NA | $\prod_{i=1}^{n}(v_{k(x_i)}/nh) = \prod_{k=1}^{K}(v_k/nh)^{v_k}$, $(0)^0 \equiv 1^*$ | See Tapi & Thompson (1990),45-46,*overlooked. 2. Depends on $t_0$, h through h & the shape. |
| UCV MISE | NA | NA | $[2 - [(n+1)/n^2]\sum_{k=1}^{K} v_k^2]/(n-1)h$ | See Scott (1992) p77, (3.52). |

**APPENDIX E  Venn diagram illustration of MOM Normal parameter estimation to compare with Fig 1.**

### Figure E Venn Diagram of MOM Normal densities

**Normal Shape (i.e. bell shaped density curves) MOM consistency etc. (Compare with Figure 1 for MOM histograms)**

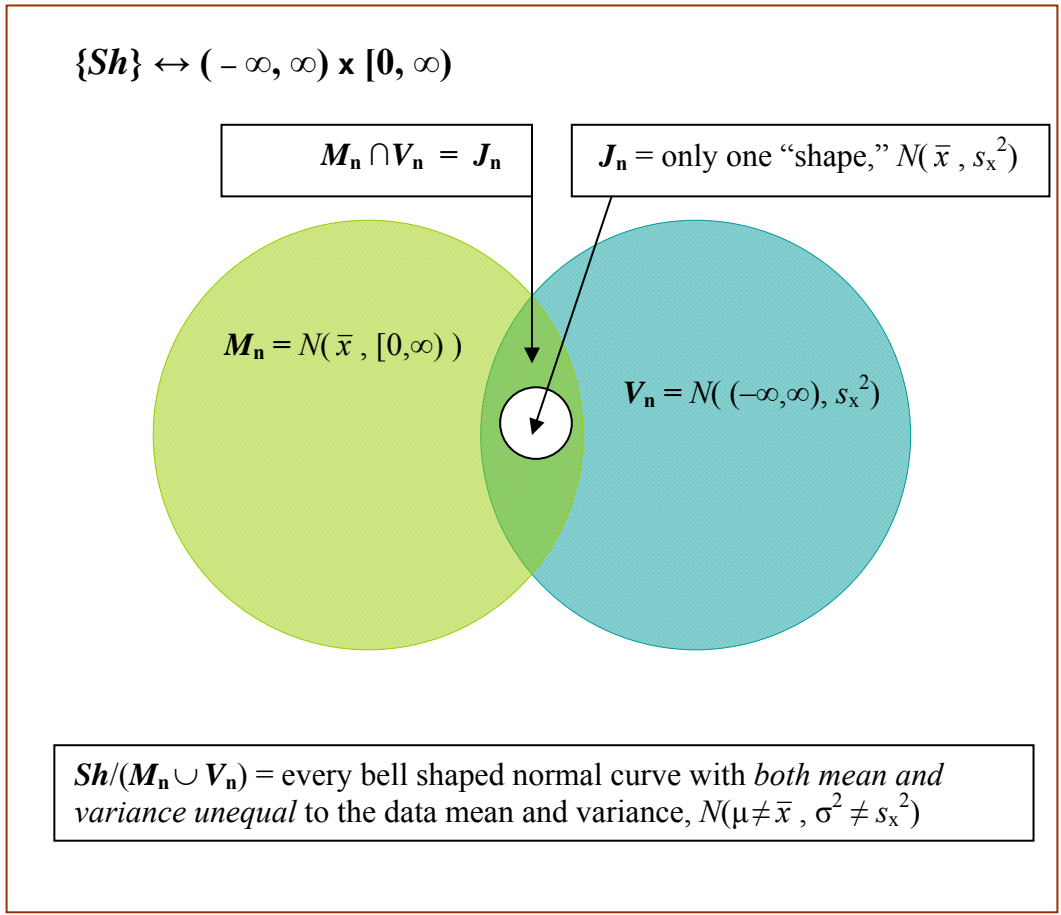

$\{Sh\} \leftrightarrow (-\infty, \infty) \times [0, \infty)$

$M_n \cap V_n = J_n$

$J_n$ = only one "shape," $N(\bar{x}, s_x^2)$

$M_n = N(\bar{x}, [0,\infty))$

$V_n = N((-\infty,\infty), s_x^2)$

$Sh/(M_n \cup V_n)$ = every bell shaped normal curve with *both mean and variance unequal* to the data mean and variance, $N(\mu \neq \bar{x}, \sigma^2 \neq s_x^2)$